\documentclass{mn2e}
\usepackage{graphicx}
\usepackage{mathptmx}
\def\SNR(#1.#2)#3(#4.#5){{G#1${\cdot}$#2$#3$#4${\cdot}$#5}}
\def\ad{{\sc ad}}
\def\OI{\mbox{O\,{\sc i}}}

\def\OIII{\mbox{O\,{\sc iii}}}
\begin{document}

\title[Far-infrared and sub-mm observations of the Crab nebula]{Far-infrared
and sub-mm observations of the Crab nebula\thanks{Based on observations with
ISO, an ESA project with instruments funded by ESA Member States (especially
the PI countries: France, Germany, the Netherlands and the United Kingdom) and
with the participation of ISAS and NASA.}}

\author[D.A.\ Green et al.]{D. A.\ Green$^1$, R. J.\ Tuffs$^2$,
                            and C. C.\ Popescu$^{2,3}$\\
   $^1$ Mullard Radio Astronomy Observatory, Cavendish Laboratory,
        Madingley Road, Cambridge CB3 0HE\\
   $^2$ Max-Planck-Institut f\"ur Kernphysik, Saupfercheckweg 1,
        D-69117 Heidelberg, Germany\\
   $^3$ The Astronomical Institute of the Romanian Academy, Str.\ Cu{\c t}itul
        de Argint 5, 752121 Bucharest, Romania}

\maketitle

\begin{abstract}
We present far-infrared (FIR) and sub-mm images of the Crab nebula, taken with
the ISOPHOT instrument on board the {\em Infrared Space Observatory} (ISO) and
with the {\em Submillimetre Common-User Bolometer Array} (SCUBA) on the James
Clerk Maxwell Telescope. The ISOPHOT observations were made in the bands
centred at 60, 100, 170 $\mu$m, with FWHM of equivalent area Gaussian beams of
44, 47 and 93 arcsec, respectively. The 850 $\mu$m SCUBA image was processed
using a Maximum Entropy Method algorithm and has a Gaussian FWHM of 17 arcsec.
The 60 and 100 $\mu$m images show clear excess of emission, above that expected
from an extrapolation of the synchrotron spectrum of the Crab nebula from lower
frequencies, as previously seen from {\sl IRAS} observations. The superior
angular resolution of the ISOPHOT images reveal that about half this excess is
attributable to two peaks, separated by $\approx 80$ arcsec. We also present
spectra taken using the {\em Long Wavelength Spectrometer} on board ISO, which
show that the FIR excess is not due to line emission. The lower resolution 170
$\mu$m image does not show any excess emission, but is possibly fainter,
particularly in the NW, than expected from an extrapolation of the
lower-frequency synchrotron emission. These findings are consistent with a
picture in which the FIR excess is due to emission from a small amount of warm
dust (in the ranges $0.01{-}0.07$ and $0.003{-}0.02$ M$_{\odot}$ for
astronomical silicate and graphite, respectively) which radiates predominantly
at 60 and 100 $\mu$m micron but not at 170 $\mu$m, and which is superimposed on
a synchrotron spectrum which gradually steepens towards shorter wavelengths
throughout the FIR and MIR spectral range. The dust geometry is consistent with
a torus of diameter $\approx 0.8$~pc created by the (red supergiant) supernova
progenitor prior to its explosion, superimposed upon a broadly distributed
component which may be supernova condensates in the filaments. The dust-to-gas
ratio in the filaments is comparable to the interstellar value. Therefore, even
if the condensates seen at the present epoch could ultimately escape the
remnant without being destroyed, the surrounding ISM will not be significantly
enriched in dust. Our upper limit of $\approx 0.02$ M$_{\odot}$ on the total
mass of Carbon in the form of graphite is consistent with the inference from
the gas-phase Carbon abundances that there has been no significant enrichment
of the filaments in Carbon nucleosynthesis products from the progenitor. To
study spectral index variations in the radio synchrotron emission we compared
the 850-$\mu$m image with a 20-cm VLA image. These images are very similar,
implying that there is little variation in spectral index across the face of
the remnant between these wavelengths. As seen previously, there are some
spectral variations near the centre of the remnant. But given the good
agreement between the integrated flux density at 850~$\mu$m and the
extrapolated synchrotron spectrum, together with the different epochs of the
850-$\mu$m and 20-cm images, we do not see the need for the second radio
synchrotron component from the remnant, which has previously been proposed.
\end{abstract}


%==============================================================================%
\section{INTRODUCTION}

The Crab nebula ($=$\SNR(184.6)-(5.8)), the remnant of the SN of {\ad} 1054
(e.g.\ Stephenson \& Green 2002), shows a centrally brightened morphology, and
it is the best known of the class of `filled-centre' supernova remnants (or
`plerions'). It is powered by its central pulsar, and emits synchrotron
emission with a relatively flat spectral index at radio wavelengths, with a
spectral index $\alpha$ here defined in the sense that flux density $S$ scales
with frequency $\nu$ as $S \propto \nu^{-\alpha}$, of $\approx 0.30$ (Baars et
al.\ 1977). The integrated spectrum of the Crab nebula steepens at higher
frequencies, with a break wavelength in the mid-infrared (e.g.\ Marsden et al.\
1984; Woltjer 1987), which is at much higher frequencies than for other
filled-centre remnants such as 3C58 ($=$\SNR(130.7)+(3.1); see Green \& Scheuer
1992), which is consistent with the central pulsar in the Crab still being
active.

\begin{table*}%----------------------------------------------------------------%
\begin{minipage}{17.5cm}
\setlength{\tabcolsep}{5pt}
\caption{Details of the ISOPHOT observations of the Crab
nebula.\label{t:isophot}}
\begin{tabular}{cccccccccccc}
 Filter & Effective  & TDT$^{\rm a}$ & \multicolumn{2}{c}{Image Centre (J2000.0)} & PA$^{\rm b}$ & Image Sampling$^{\rm c}$ & Image Size      & Background  & FWHM$^{\rm d}$ &       Integrated       \\
        & wavelength &               & RA                &             Dec        &              & $Y \times Z$           &                 &             &                & flux density$^{\rm e}$ \\
        & ($\mu$m)   &      & (h\quad m\quad s) & ($^\circ$\quad$'$\quad $''$)    & (degree)     & (arcsec$^2$)           & (arcsec$^2$)    &  (MJy/sr)   &  (arcsec)      &         (Jy)           \\ \hline
   C60  &     60     &    82301720   &     05 34 32.44   &     $+22$ 01 14.5      &     356.80   &   $15.32\times23.00$   & $570\times 520$ &    25.6     &     44         &      $140.7 \pm 11\%$  \\
  C100  &    100     &    82301720   &     05 34 32.44   &     $+22$ 01 14.5      &     356.81   &   $15.32\times23.00$   & $570\times 520$ &    28.9     &     47         &      $128.2 \pm 11\%$  \\
  C160  &    170     &    82401805   &     05 34 31.70   &     $+22$ 00 48.2      &     356.84   &   $30.65\times46.00$   & $766\times 506$ &    52.0     &     93         &      $ 83.2 \pm 12\%$  \\ \hline
\end{tabular}

\medskip\noindent Notes:\\
$^{\rm a}$ Target Dedicated Time identifier. The first three digits give the
orbit identifier, which is also the epoch of the observation in days after 1995
November 17th.\\
$^{\rm b}$ Positive Y direction (the direction of the chopper sweep), degrees E
from N.\\
$^{\rm c}$ Spacecraft coordinates.\\
$^{\rm d}$ FWHM of equivalent area circular Gaussian.\\
$^{\rm e}$ The uncertainties are a combination of those due to the uncertainty
in the spectrum of the absolute calibrators (10 per cent in each band) and long
term variations in detector responsivity (determined from the dispersion in the
response of individual pixels to the background to be 3 per cent at 60 and 100
${\mu}$m and 6 per cent at 170 ${\mu}$m). The systematic uncertainties
dominate the random uncertainties, which at 60 and 100 ${\mu}$m are mainly due
to residual glitches in the detector signal response -- at a level of $\approx
1$ per cent --  and at 170 ${\mu}$m are mainly due to the shot noise induced by
the illumination of the target and background -- at a level of $\approx 2$ per
cent.\\
\end{minipage}
\end{table*}%------------------------------------------------------------------%

In addition to the evidence for a break in the mid-infrared synchrotron
spectrum, IRAS observations (Marsden et al.) also revealed the existence of a
far-infrared (FIR) excess in the integrated measurements over the extrapolation
of the radio--sub-mm synchrotron spectrum. It has been suggested that this
excess could have one or a mixture of two distinct physical origins -- warm
($\sim 45~K$) dust emission (Marsden et al.\ 1984; Mezger et al.\ 1986; Strom
\& Greidanus 1992) and synchrotron emission arising from a bump in the
relativistic electron energy spectrum around the energy corresponding to the
synchrotron break wavelength (Mezger et al.\ 1986). Furthermore, Fesen \& Blair
(1990) suggested the existence of grains in the nebula on the basis of optical
measurements. Their study of various optical images revealed numerous dark
spots, ranging up to 5 arcsec in size, which they attributed to patches of
obscuration. Evidence was also found for an extended component of the
obscuration. However the attributes of prime significance -- the mass,
composition and temperature of the grains -- could not be addressed by this
investigation, nor by the subsequent analysis using images from the Hubble
Space Telescope (HST) by Blair et al.\ (1997). In a separate study, Fesen,
Martin \& Shull (1990) made a case that the bays seen prominently in optical
continuum images at the E and W extremities of the remnant represent the
projection of a toroid of pre-existing material created by a (red giant)
progenitor wind (see also Li \& Begelman 1992). They supposed that this disk
could account for some of the high Helium abundances seen in the remnant and
could be dusty. Douvion et al.\ (2001) presented some infrared observations of
the Crab nebula, made with the ISOCAM instrument aboard the ISO satellite.
These observations, which were made at 6 to 16 $\mu$m (i.e.\ wavelengths just
shortwards the mid-infrared break), did not detect any dust emission, but only
synchrotron emission, which showed a steepening of the mid-infrared spectral
index away from the central pulsar. It is important to search for cold grains
at longer wavelengths, since it not known whether significant quantities of
dust are produced by the progenitor star of the type that produced the Crab
nebula, either in a stellar wind prior to the outburst or as condensates in the
metal rich supernova ejecta. From observed elemental abundances, Nomoto (1984)
concludes that the progenitor of the Crab nebula must have been $\la
13$~M$_\odot$, with a mass of $\sim 9$~M$_\odot$ being preferred. Recent sub-mm
observations (Dunne et al.\ 2003; Morgan et al.\ 2003) suggest that there are
large amounts of cold dust -- of the order of a Solar mass -- in the supernova
remnants Cassiopeia A and that of Kepler's supernova of {\ad} 1604 (although an
alternative explanation for the sub-mm emission from Cas A has been proposed by
Dwek 2004). Sources of dust production are important for understanding dust
high-redshift galaxies (e.g.\ Morgan \& Edmunds 2003), and in this context
further investigations of the dust content of Galactic supernova remnants are
useful.

Concerning the synchrotron emission, although there have been claims of
variations in spectral index across the Crab nebula at radio wavelengths (e.g.\
between the filaments and the diffuse inter-filament regions, and a systematic
steepening of the spectrum towards the edge of the remnant), these were not
confirmed by the detailed study of Bietenholz et al.\ (1997). The spectral
variations that were detected were: (i) within about an arcminute of the
pulsar, due to moving features in the radio emission (see also Bietenholz,
Frail \& Hester 2001; Bietenholz et al.\ 2004), and the different epochs of the
images compared, and (ii) some absorption at low frequencies (below about
300~MHz), due to thermal material in the Crab nebula's filaments. More recent
comparisons by Bietenholz, et al.\ (2001) show that radio observations of the
Crab nebula taken about a year and half apart indeed show changes in the
structure of the emission from the Crab nebula close to its central pulsar.
However, any spectral variations are difficult to detect over a narrow range of
wavelengths, and may be more easily detectable if a good mm or sub-mm image of
the Crab nebula is available for comparison with longer wavelength radio
observations. Bandiera, Neri \& Cesaroni et al.\ (2002) presented a 240-GHz
(i.e.\ 1.3~mm) image of the Crab nebula, and compared it with a 20-cm image for
spectral index studies, from which they proposed a second synchrotron component
to the emission from the remnant.

In this paper we present images of the nebula in both the FIR (ISOPHOT images
at 60, 100 and 170~$\mu$m) and the sub-mm (a SCUBA image at 850~$\mu$m), as
well as four FIR spectra taken with the LWS instrument on board ISO. (A
preliminary analysis of the sub-mm observations was presented by Green 2002.)
These allow us to spatially and spectrophotometrically distinguish between
synchrotron, dust and line emission, to determine the amount and distribution
of cold dust in the nebula, and to investigate further the radio to sub-mm
spectral index across the remnant. J2000.0 coordinates are used throughout this
paper, and we take the distance to the Crab nebula to be $\approx 2$~kpc
(Trimble 1973). The FIR and sub-mm observations are described in Section
\ref{s:iso-obs} and \ref{s:jcmt-obs} respectively. These observations are
discussed is \ref{s:discussion}, and our conclusions summarised in Section
\ref{s:conclusions}.

\begin{figure*}%---------------------------------------------------------------%
\centerline{\includegraphics[width=7.5cm]{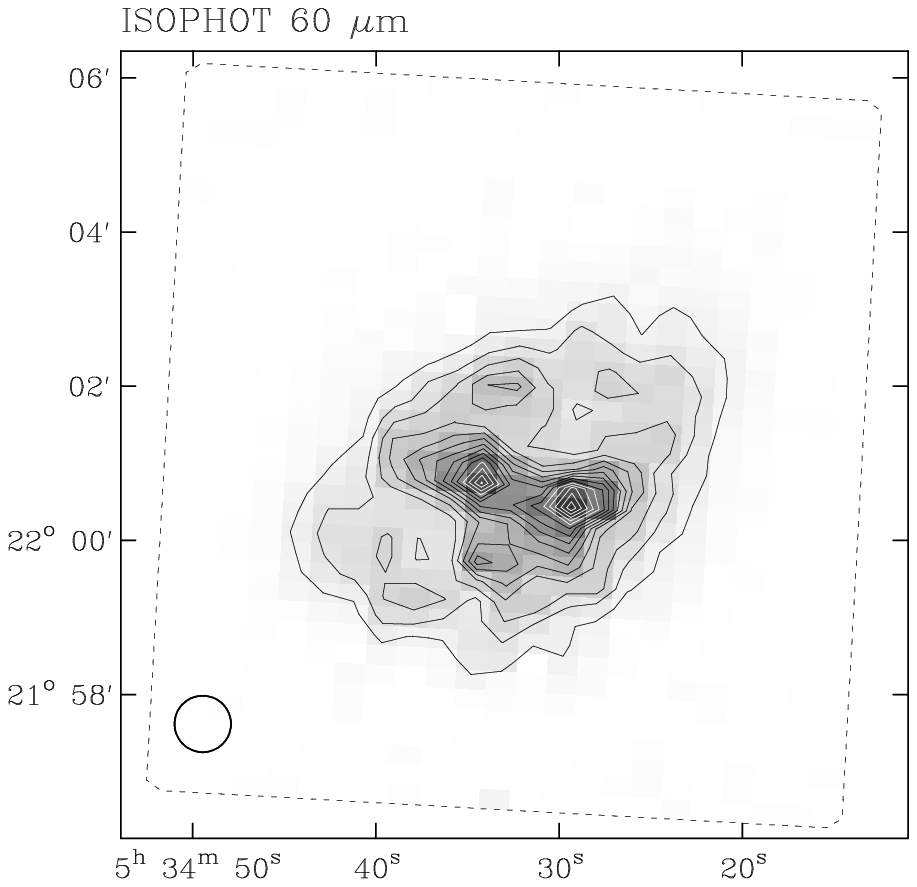} \quad
            \includegraphics[width=7.5cm]{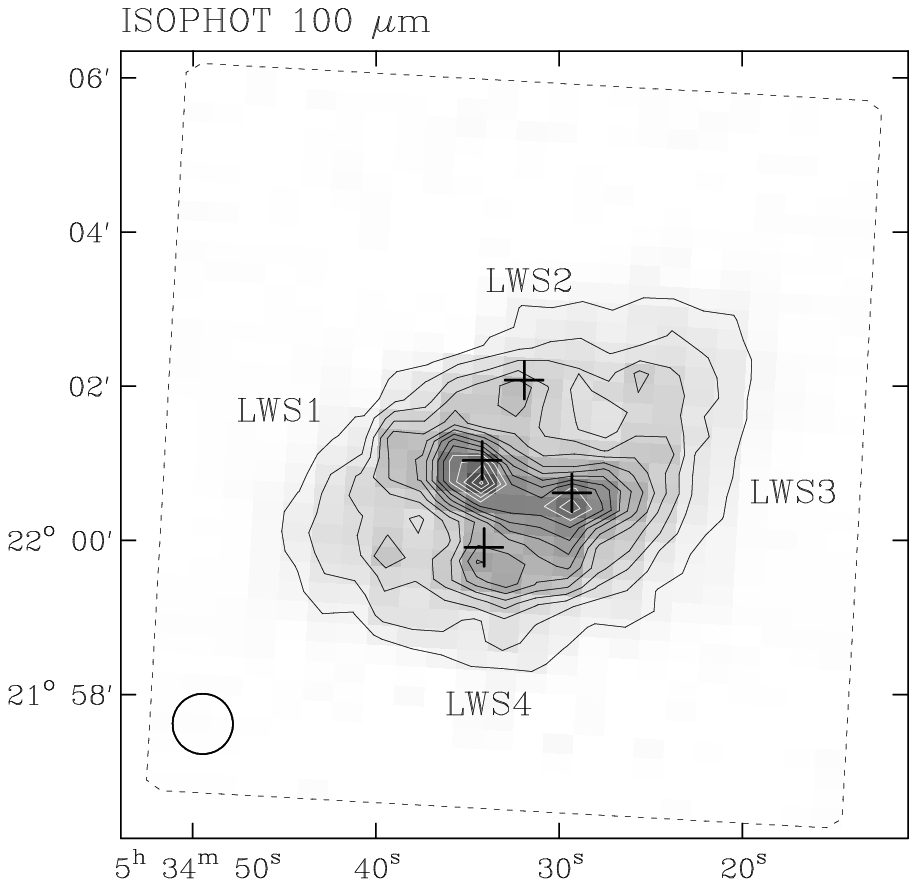}}
\caption{ISOPHOT images of the Crab nebula at 60 and 100~$\mu$m, with
resolutions of 43.5 and 46.8 arcsec respectively. The contours are every 25 MJy
sr$^{-1}$ at 60 $\mu$m, and every 18 MJy sr$^{-1}$ at 100 $\mu$m, and the
dashed rectangles indicate the regions observed. The circles in the lower left
indicate the resolution of the observations. The labelled crosses on 100 $\mu$m
image indicate where LWS spectra were obtained (see Section~\ref{s:iso-lws} and
Table~\ref{t:isolws}).\label{f:iso60and100}}
\end{figure*}%-----------------------------------------------------------------%

%==============================================================================%
\section{ISO DATA REDUCTION AND RESULTS}\label{s:iso-obs}

\subsection{ISOPHOT images}\label{s:iso-isophot}

\begin{figure}%----------------------------------------------------------------%
\centerline{\includegraphics[width=8.0cm]{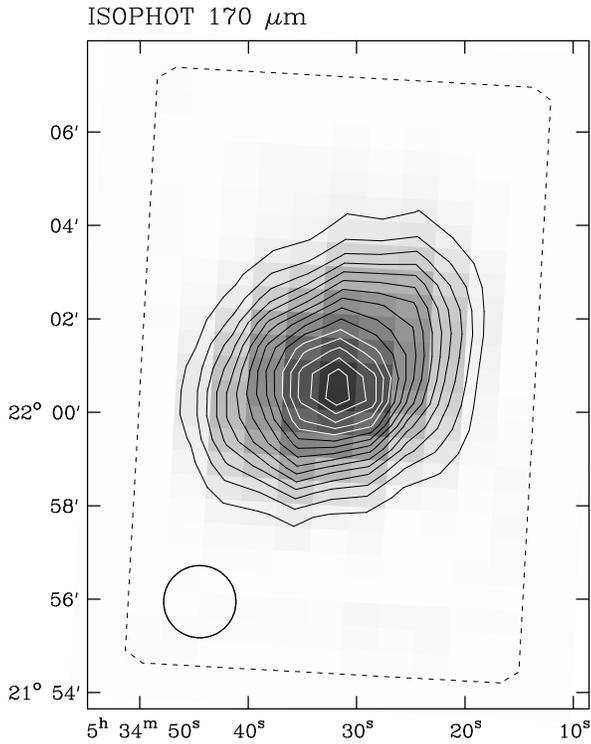}}
\caption{ISOPHOT images of the Crab nebula at 170 $\mu$m, with a resolution of
92.6 arcsec. The contours are every 5 MJy sr$^{-1}$, and the dashed rectangle
indicates the region observed. The circle in the lower left
indicates the resolution of the observations.\label{f:iso160}}
\end{figure}

Details of the FIR observations of the Crab nebula made with the ISOPHOT
instrument of the {\sl Infra-Red Space Observatory} (ISO; see Kessler et al.\
1996 for details about ISO, and Lemke et al.\ 1996 for details of ISOPHOT) are
given in Table~\ref{t:isophot}. These were made using the ISOPHOT-C100
3$\times3$ pixel array with the C60 and C100 filters and the ISOPHOT-C200
2$\times2$ pixel array with the C160 filter. The passbands of the C60, C100 and
C160 filters are between 48--73 $\mu$m, 82--124 $\mu$m and 130--218 $\mu$m, and
have effective central wavelengths near 60, 100 and 170 ${\mu}$m, respectively.
(Note that the name of the C160 filter does not correspond to the central
wavelength, unlike the case of the C60 and C100 filters.) The `P32' mapping
mode was used to provide near Nyquist sampling over an area encompassing the
nebula, as well as the surrounding background. The total exposure was 50
minutes, corresponding to exposures per image pixel exposure of 4.7 s at both
60 and 100 $\mu$m, and 18.8 s at 170 $\mu$m. The area of the beam covers about
6 independent pixels in each filter, so that the effective exposure time per
beam is about six times longer in each case. The data for each wavelength were
separately processed using the latest P32 reduction package (Tuffs \& Gabriel
2003), which corrects for the transient response of the detector pixels. A
time-dependent flat-field correction was made for each image, by fitting a
quadratic function to the response of the detector pixels to the background.
Calibration was made using V8.1 of the ISOPHOT Interactive Analysis (PIA)
Package (Gabriel et al.\ 1997), which is based on the on-board calibration
source (which was periodically checked with respect to the primary celestial
calibrators). Although the images are oversampled, independent data contribute
to each image pixel. Finally, the background was removed by subtracting a
tilted plane obtained from a fit to the extremities of the image (external to
the extent of the nebula). The derived images at 60, 100 and 170 ${\mu}$m are
shown in Figs~\ref{f:iso60and100} and \ref{f:iso160}.

%------------------------------------------------------------------%

\begin{table*}%----------------------------------------------------------------%
\begin{minipage}{15.0cm}
\caption{Details of LWS spectra observed towards the Crab nebula, and the
measured contributions of the detected line emission to the ISOPHOT and IRAS 60
and 100 $\mu$m images.\label{t:isolws}}
\tabcolsep3pt
\begin{tabular}{cccclcccccc} \hline
Label &    Position   & \multicolumn{2}{c}{PHT measured$^{\rm a}$} & \quad line & line power              & \multicolumn{4}{c}{Equivalent in-filter continuum brightness$^{\rm b}$} \\      % & LWS$^{\rm c}$    %   PHT pixel
      &    (J2000.0)  &        C60      &      C100      &                      & ($10^{-14}$ W m$^{-2}$) &     PHT-C60  &   PHT-C100  & IRAS-60 $\mu$m & IRAS-100 $\mu$m \\                % & (MJy/sr)         %     (Y, Z)
      &               &     (MJy/sr)    &    (MJy/sr)    &                      &                         &    (MJy/sr)  &   (MJy/sr)  &    (MJy/sr)    &   (MJy/sr)      \\ \hline         % &                  %
%                                                                                                                                                                                           %                    %
 LWS1 &   05 34 34.2  &         195     &        181     & [\OIII] 52 $\mu$m    & 4.1                     &        8     &      --     &         7      &       --        \\                % &     784          %  10.0, 18.9
      &  +22 01 02.3  &                 &                & [\OI]   63 $\mu$m    & 1.2                     &        5     &      --     &         3      &       --        \\                % &     336          %
      &               &                 &                & [\OIII] 88 $\mu$m    & 3.2                     &       --     &      13     &        --      &       13        \\ [2pt]          % &    1752          %
      &               &                 &                & Total                &                         &       13     &      13     &        10      &       13        \\ [4pt]          % &                  %
%                                                                                                                                                                                           %                    %
 LWS2 &   05 34 31.9  &          90     &        76      & [\OIII] 52 $\mu$m    & 2.1                     &        4     &      --     &         4      &       --        \\                % &     401          %  14.8, 11.1
      &  +22 02 04.8  &                 &                & [\OI]   63 $\mu$m    & 1.1                     &        4     &      --     &         3      &       --        \\                % &     308          %
      &               &                 &                & [\OIII] 88 $\mu$m    & 1.7                     &       --     &       2     &        --      &        3        \\ [2pt]          % &     931          %
      &               &                 &                & Total                &                         &        8     &       2     &         7      &        3        \\ [4pt]          % &                  %
%                                                                                                                                                                                           %                    %
 LWS3 &   05 34 29.3  &         216     &        161     & [\OIII] 52 $\mu$m    & 7.1                     &       13     &      --     &        12      &       --        \\                % &    1356          %  20.4, 13.0
      &  +22 00 37.0  &                 &                & [\OI]   63 $\mu$m    & 2.0                     &        8     &      --     &         5      &       --        \\                % &     561          %
      &               &                 &                & [\OIII] 88 $\mu$m    & 4.4                     &       --     &      17     &        --      &       18        \\ [2pt]          % &    2245          %
      &               &                 &                & Total                &                         &       21     &      17     &        17      &       18        \\ [4pt]          % &                  %
%                                                                                                                                                                                           %                    %
 LWS4 &   05 34 34.1  &         132     &        121     & [\OIII] 52 $\mu$m    & 2.7                     &        5     &      --     &         5      &       --        \\                % &     516          %  23.4, 10.2
      &  +21 59 54.6  &                 &                & [\OI]   63 $\mu$m    & 1.5                     &        6     &      --     &         4      &       --        \\                % &     421          %
      &               &                 &                & [\OIII] 88 $\mu$m    & 1.9                     &       --     &       8     &        --      &        8        \\ [2pt]          % &     969          %
      &               &                 &                & Total                &                         &       11     &       8     &         9      &        8        \\ \hline         % &                  %
\end{tabular}

\medskip\noindent Notes:\\
$^{\rm a}$ after convolution to a Gaussian approximation
(FWHM 68 arcsec) to the LWS beams measured by Lloyd (2003).
$^{\rm b}$ The equivalent brightness of the diffuse continuum emission which
would give rise to the same power received by the ISOPHOT or IRAS detectors as
received by the LWS from the emission lines.\\
\end{minipage}
\end{table*}%------------------------------------------------------------------%

\begin{figure*}%---------------------------------------------------------------%
\centerline{\includegraphics[width=16cm]{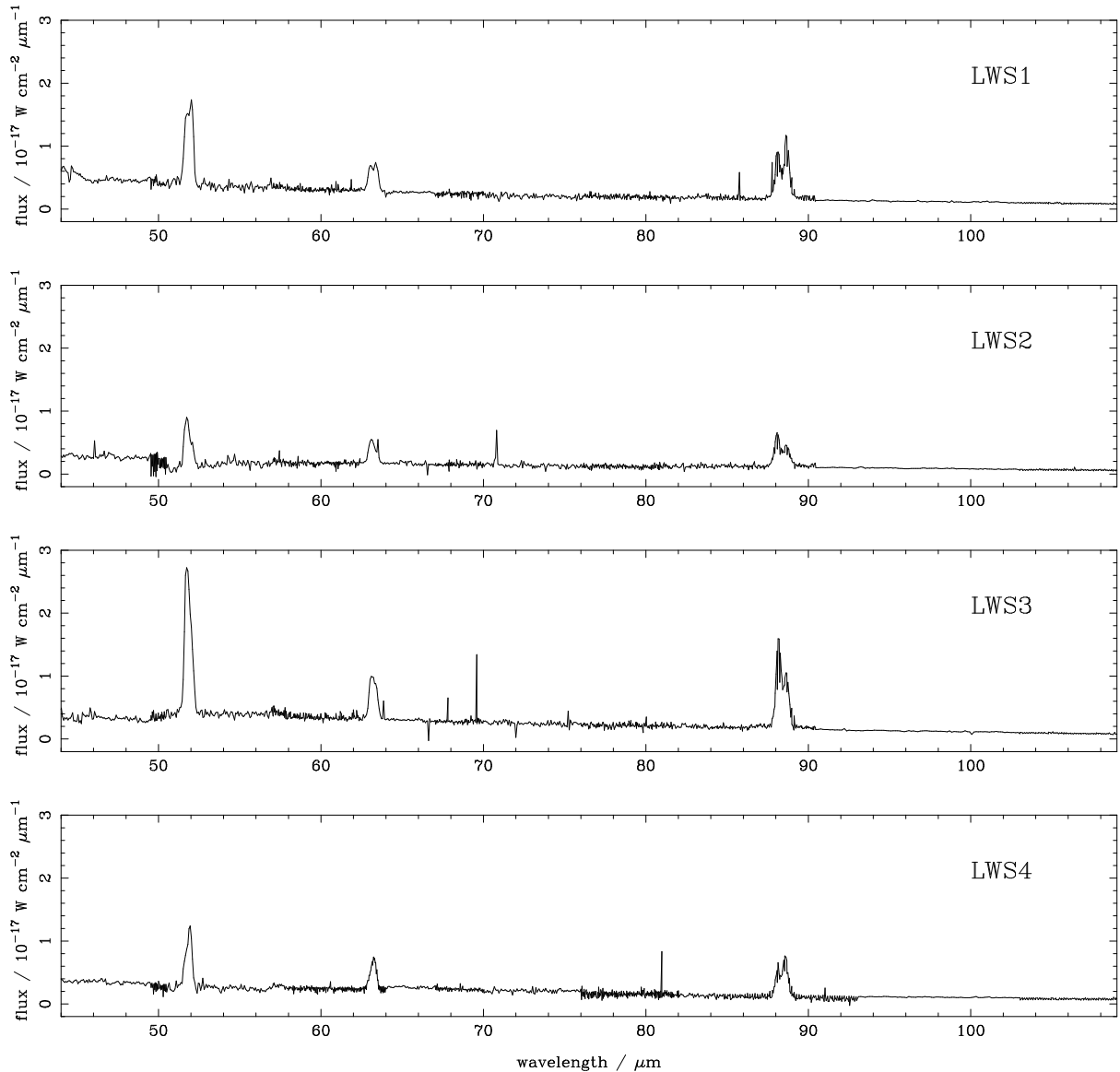}}
\caption{LWS spectrum from four positions towards the Crab nebula, at points
indicated on the 100 $\mu$m image in Fig.~\ref{f:iso60and100}.\label{f:isolws}}
\end{figure*}%-----------------------------------------------------------------%

The integrated flux densities of the nebula at the 60, 100 and 170 ${\mu}$m,
together with the backgrounds, are given in Table~\ref{t:isophot}. The ISO
backgrounds are consistent with the backgrounds measured from COBE/DIRBE to
within about 15 per cent in C160 and to within about 10 per cent at 100 and at
60 $\mu$m. These accuracies are limited by the background dust emission from
the Milky Way in the COBE 40 arcmin beams. Even though the ISOPHOT data have
been corrected for signal losses due to the transient response of the
detectors, the integrated flux densities at 60 and 100 $\mu$m are lower by
about 30 per cent compared with those found by IRAS in the corresponding bands.
This seems to be a general phenomenon, and not just restricted to datasets
considered in this paper; the discrepancies seem to be larger for extended
sources than for point-like sources (see Tuffs \& Gabriel 2003; Tuffs et al.\
2002). This difference between ISO and IRAS photometry of extended sources may
be due to the fact that a single correction factor is used to correct for
signal losses due to the transient response of the IRAS detectors (see the IRAS
Explanatory Supplement, Beichman et al.\ 1988), whereas the corrections for
the ISOPHOT data self-consistently take into account the source structure.
Since the IRAS correction factor was derived from measurements of point source
primary calibrators, and it is inherent to the transient response of the Ge:Ga
detectors that the shortfall in signal will be greater for point sources than
for extended sources, this may explain the higher integrated flux densities
obtained for IRAS observations of extended sources in the 60 and 100 $\mu$m
bands compared with ISOPHOT.

\subsection{LWS spectra}\label{s:iso-lws}

Spectroscopic observations were made towards four directions in the nebula
(marked by crosses superimposed on the 100 $\mu$m image of
Fig.~\ref{f:iso60and100}) with the Long Wavelength Spectrometer (LWS)
instrument onboard ISO (see Clegg et al.\ 1996). These spectra show the
presence of strong emission from all the [\OI] and [\OIII] lines. Details of
these spectra, and the results, are given in Table~\ref{t:isolws}, and the
spectra are shown in Fig.~\ref{f:isolws}. The spectra are standard pipeline
products (OLP Version 10), taken from the ISO archive (from a program of D.\
P\'equignot), and include some spikes due to cosmic rays. Two of the positions
observed  (LWS1 and LWS3) were close to the two FIR peaks seen on the ISOPHOT
60 and 100 $\mu$m image, near to the centre of the Crab nebula (see
Fig.~\ref{f:iso60and100}), with the other two positions nearer the edge of the
nebula.

%==============================================================================%
\section{JCMT DATA REDUCTION AND RESULTS}\label{s:jcmt-obs}

The Crab nebula was observed with the {\em Submillimetre Common-User Bolometer
Array} (SCUBA) (Holland et al.\ 1999) using the `850 $\mu$m' filter on the
James Clerk Maxwell Telescope (JCMT) on 1999 September 19. SCUBA operates
at both 850 and 450 $\mu$m simultaneously, but since the surface of the
telescope was not well set when the observations were made, only the results at
850 $\mu$m are presented. The 850 $\mu$m SCUBA filter is actually centred at
863 $\mu$m (i.e.\ 347~GHz). At this wavelength SCUBA has 37 bolometers, each
with an ideal resolution of 13~arcsec, covering a field-of-view of $\approx 2$
arcmin. Since the Crab nebula is significantly larger than the SCUBA
field-of-view, the observations were made in the {\tt scan-map} mode, where the
array scans across the source with the telescope continuously `chopping' in a
particular direction. A variety of chop-throws and scanning directions were
observed, in order to sample structure well in all directions, and on scales
missed by any single chop throw. In total six chop throws were used (30, 44 and
68 arcsec in both RA and DEC), each for three different scanning directions (at
position angles, PAs, of $15\fdg5$, $75\fdg5$ and $135\fdg5$). A region
$9\times7$ arcmin$^2$, at a PA of $45^\circ$, was observed in order to ensure a
clear emission-free region around the Crab nebula was covered. The observations
were made in two sessions, over about 4.5 hours, at elevations between $\approx
60^\circ$ and $\approx 80^\circ$. Each session was preceded by observations of
the standard source CRL 618, and was preceded and followed by a sky-dip
calibration. The observing conditions varied little, as indicated by both the
sky-dip observations, and the Caltech Submillimeter Observatory `tau-meter'
readings at 225~GHz (which varied between 0.048 and 0.065 for the Crab nebula
observations).

\begin{figure}%---------------------------------------------------------------%
\centerline{\includegraphics[width=8.5cm]{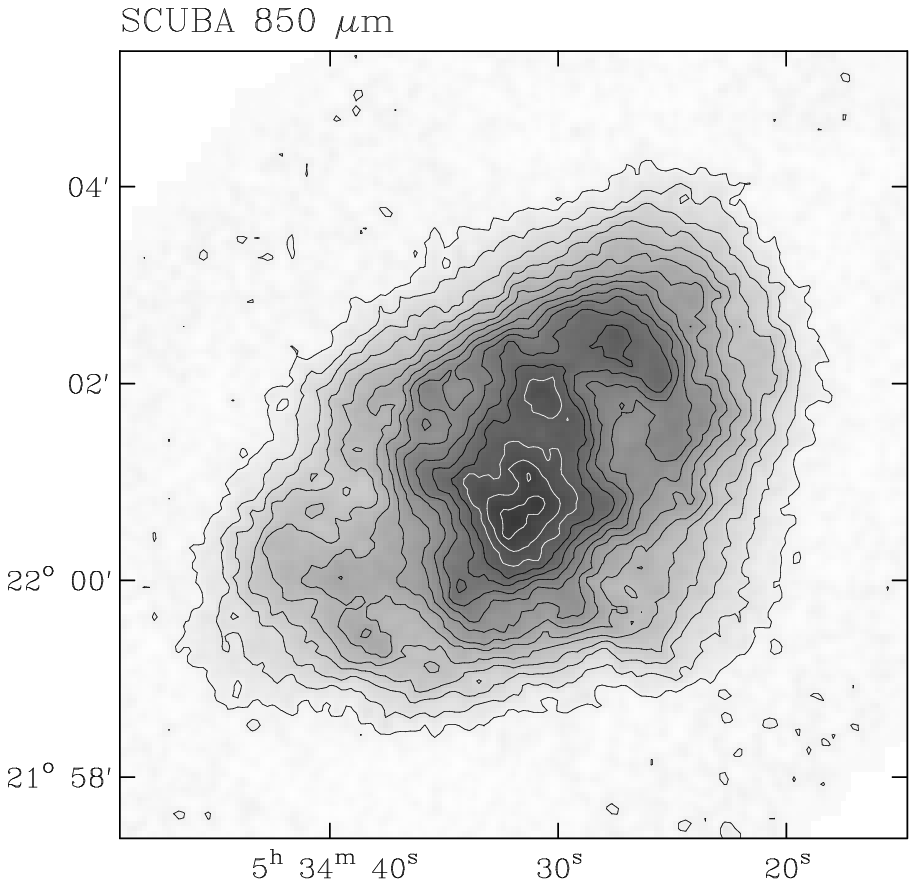}}
\caption{The Crab nebula at 850~$\mu$m (347~GHz) from the JCMT SCUBA
observations, with a resolution of 17~arcsec. Contours are every 0.08
Jy~beam$^{-1}$.\label{f:850}}
\end{figure}%-----------------------------------------------------------------%

\begin{figure*}%---------------------------------------------------------------%
\centerline{\includegraphics[angle=270,width=17cm]{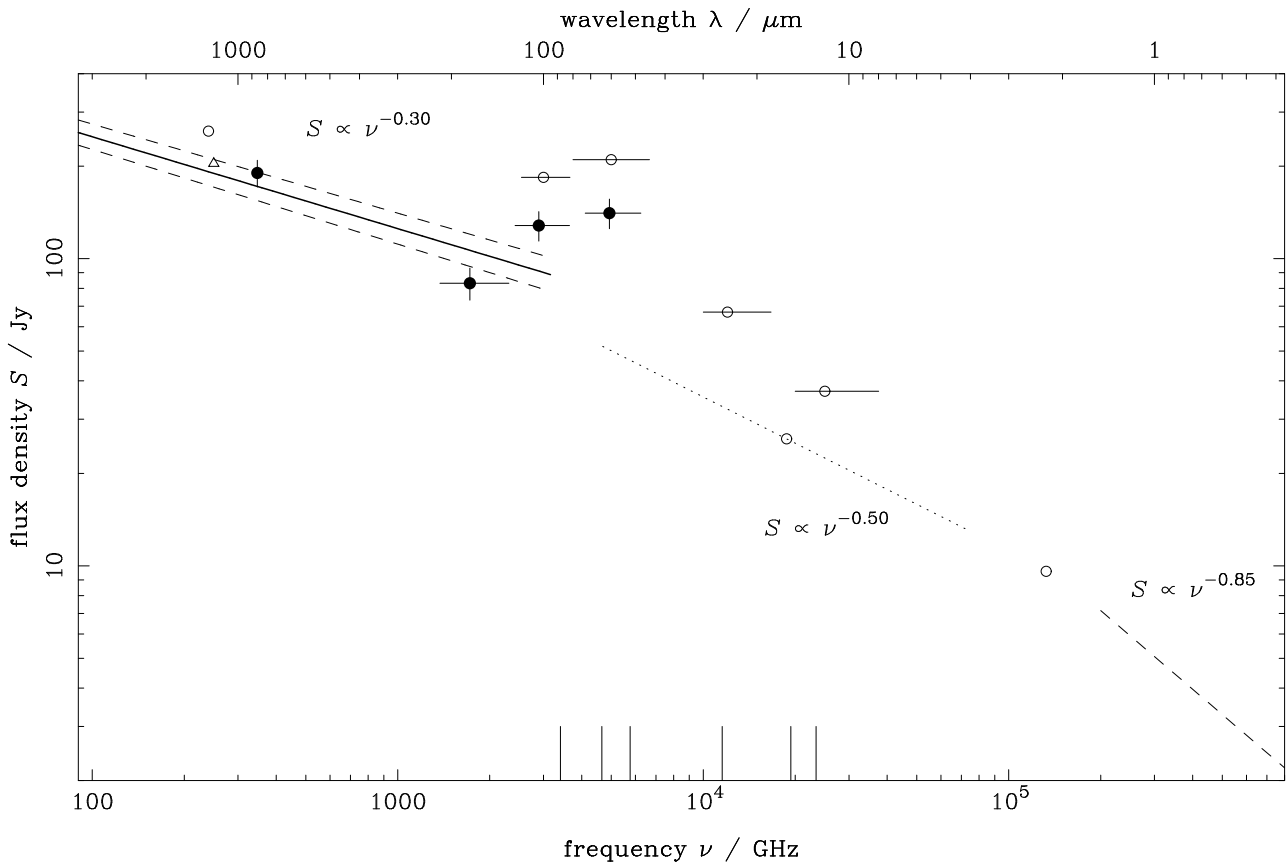}}
\caption{The sub-mm/infra-red spectrum of the Crab nebula. The filled circles
are from the JCMT (850 $\mu$m) and ISO (170, 100 and 60 $\mu$m) observations
reported here. The open circles are flux densities: (a) at 1.2-mm from Bandiera
et al.\ (2002); (b) from IRAS at 100, 60, 25 and 12 $\mu$m from Strom \&
Greidanus (1992) -- see Section~\ref{s:iso-isophot} for a discussion of the
discrepancy between the IRAS and ISO photometry at 60 and 100 $\mu$m; (c) a
scaled pure continuum flux density at 16 ${\mu}$m from Douvion et al.\ (2001)
-- see text for details of the scaling -- together with a pure continuum
spectrum through this point with $\alpha=0.50$ (dotted line); (d) a pure
continuum flux density at 2.26~$\mu$m -- see text for details. The open
triangle is an additional 1.2-mm flux density from Mezger et al.\ (1986). The
range of frequencies covered is indicated for the broad band IRAS and ISO
observations. The solid line at lower frequencies is an extrapolation of the
synchrotron spectrum from radio frequencies from Baars et al.\ (1977), with
errors shown as dashed lines. The dashed line at higher frequencies is an
extrapolation of the $S = 12.9 (\nu / 10^5 {\rm\ GHz})^{-0.85}$~Jy optical/UV
spectrum from Woltjer (1987). The vertical lines at the bottom of the plot mark
the positions of lines at 12.8, 15.5, 26, 52, 63 and 88
$\mu$m.\label{f:spectrum}}
\end{figure*}%-----------------------------------------------------------------%

The data were reduced using a series of standard procedures from the {\em SCUBA
User Reduction Facility} (SURF) package (see Jenness \& Lightfoot 2000). This
processing included corrections for the extinction at 850~$\mu$m, as measured
by sky-dip observations (the observed optical depths varied between 0.19 and
0.21); removal of spikes in the data, both manually and automatically; removal
of six poorly-performing bolometers (four of which were at the edge of the
bolometer array); removal of linear baselines; and removal of sky
contributions. Finally the data were restored to an image using a MEM algorithm
(Pierce-Price 2001).

The flux density scale was set by the observations of the calibrator source CRL
618 (with an assumed flux density of 4.56~Jy at 850 $\mu$m, from the JCMT
calibrator webpages,\footnote{See: {\tt http://www.jach.hawaii.edu/}} see also
Sandell 1994). The scaling factor used was the average of those determined from
the two available observations of CRL 618, which differed by only $\pm 2$
per~cent from their mean (with this flux scale, the integrated flux density of
the Crab nebula is in good agreement with that expected from extrapolation of
data from lower frequencies, see below). From the CRL 618 observations the beam
was fitted with a Gaussian of HPBW of 16 arcsec.

Figure~\ref{f:850} shows the emission from the Crab nebula at 850 $\mu$m from
these observations, smoothed slightly to a resolution of 17 arcsec. The noise
on this image, in small regions away from the Crab nebula, is $\approx 0.017 $
Jy~beam$^{-1}$, with variations in the local baselevel away from the Crab
nebula, up to $\approx 0.05 $ Jy~beam$^{-1}$. This image has considerably
higher dynamic range than the 1.3 mm image of Bandiera et al., although it does
have slightly poorer resolution.

The integrated flux density of the Crab nebula from the SCUBA 850-$\mu$m
observations is $\approx 190$~Jy. This was determined by integrating
within polygons drawn beyond to boundary of emission from the remnant. The
flux density obtained varied by up to $\approx 2$ per cent, depending on where
an individual polygon was drawn. We conservatively take the uncertainty in the
integrated flux density to be 10 per cent, to include the uncertainties in
choice of polygon to integrate over, the small change in amplitude scaling seen
from the observations of CRL 618 above, and any uncertainty in the assumed flux
density for CRL 618. As discussed below, this is in good agreement with the
prediction of the expected flux density from the extrapolation of the
synchrotron spectrum of the Crab nebula at lower frequencies.

%==============================================================================%
\section{DISCUSSION}\label{s:discussion}

\subsection{Sub-mm/Infrared Spectrum}\label{s:spectrum}

The integrated spectrum of the Crab Nebula is shown in Fig.~\ref{f:spectrum},
together with an extrapolation of the synchrotron spectrum into the infrared
domain from the UV/optical domain (from Woltjer 1987) and from the radio domain
(from Baars et al.\ 1977). In making the extrapolation from the radio domain we
took into account the fact Baars et al.\ give a fitted spectrum for the
synchrotron emission from the Crab nebula for frequencies up to 35~GHz, from
observations predominantly made in the late 1960s. Since the Crab nebula is
known to be fading at a rate of $0.167 \pm 0.015$ per cent year$^{-1}$ at 8~GHz
(Aller \& Reynolds 1985), the extrapolated spectrum shown in
Fig.~\ref{f:spectrum} has been decreased by 5.5 per cent (i.e.\ corresponding
to a period of  33 years, the difference mean date of the observations used by
Baars et al.\ and our observations). The uncertainties in the extrapolation of
the Baars et al.\ spectrum are: (i) the overall uncertainty in amplitude scale
(which is $10^{\pm 0.031}$, i.e.\ $\approx 7.4$ per cent); (ii) a frequency
dependent uncertainty due to the uncertainty in the spectral index (which is
$(\nu/\nu_0)^{\pm 0.009}$); (iii) any uncertainty if the fading at higher
frequencies is different from that measured by Aller \& Reynolds at 8~GHz,
which is as yet unknown. The uncertainty in the spectral index -- using a
reference frequency of 7~GHz, which is close to the centre of the frequencies
of the observations used by Baars et al.\ -- corresponds to uncertainties of
$\approx 3.5$ per cent at 850~$\mu$m, increasing to $\approx 6.2$ per cent at
60~$\mu$m. Adding the uncertainties due to the overall amplitude scale of the
fitted spectrum and the spectral index in quadrature gives an uncertainty in
the extrapolated spectrum of $\approx 8.2$ per cent at 850~$\mu$m, increasing
to $\approx 9.7$ per cent at 60~$\mu$m, as is indicated by the dashed lines at
$\lambda < 100$ ${\mu}$m in Fig.~\ref{f:spectrum}. We note the following.
\begin{enumerate}
\item The ISOPHOT 170 $\mu$m flux density is somewhat low compared with the
extrapolated Baars et al.\ spectrum, although this is not highly significant
(see further discussion below).
\item Both the 100 and 60 $\mu$m flux densities show clear evidence for an
excess above the extrapolated synchrotron spectrum (see
Section~\ref{s:isoimages} for further discussion), albeit at a somewhat lower
level than has been seen previously from {\sl IRAS} observations (e.g.\ Strom
\& Greidanus). This excess covers a range of frequencies that is too narrow for
it to be synchrotron emission. Further evidence that the FIR excess is thermal
is provided by the polarisation measurements of Klaas et al.\ (1999), which
show that the excess emission is unpolarised.
\item The 850-$\mu$m flux density is consistent with the extrapolated Baars et
al.\ spectrum, unlike the 1.2-mm flux density from Bandiera et al.\ (see
further discussion in Section~\ref{s:cmtosubmm}).
\end{enumerate}
The expected flux density from the extrapolation of the Baars et al.\ spectrum
at 170 $\mu$m is $105.5 \pm 9.5$ Jy, whereas the 170 ${\mu}$m integrated flux
density is $83.2 \pm 9.8$ Jy (see Table~\ref{t:isophot}). The difference in
these values is $22 \pm 14$ Jy, which is not highly significant. If this
deficit is real, it could be indicative of spectral turnover between 850 and
170~$\mu$m. Further evidence for a steepening of the FIR synchrotron spectrum
at these wavelengths is provided by the differences in the surface brightness
distributions at 850 and 170 $\mu$m (see further discussion in section
\ref{s:170and850}). Overall, our data are more consistent with a gradual
transition between the $\nu^{-0.3}$ radio spectrum and the $\nu^{-0.85}$
UV--optical spectrum (Woltjer 1987) spanning the whole of the FIR--near-IR,
rather than a sharp break in the mid-IR. Support for a gradual turnover in the
infrared synchrotron emission is provided by the ISOCAM CVF measurements of
Douvion et al.\ (2001), which had sufficient spectral resolution to separate
the continuum and line components. Douvion et al.\ found that the continuum
emission between 6.5 and 16 $\mu$m is well described by a $\nu^{-0.5}$ power
law -- i.e.\ with a slope intermediate between the slopes of the radio and the
UV--optical power laws. We have also plotted the MIR continuum spectrum of
Douvion et al.\ in Fig.~\ref{f:spectrum}, multiplied by a correction factor of
1.29 to compensate for the fact that Douvion et al.\ did not image the whole of
the Crab nebula. This correction factor was determined from a image of the 2.26
$\mu$m continuum emission of the nebula by measuring the ratio of the flux
density integrated over the whole source to the flux density integrated over a
synthetic $192 \times 192$ arcsec$^2$ aperture representing the ISOCAM field of
view. Also shown in Fig.~\ref{f:spectrum} is the integrated continuum flux
density found from an image (Gallant \& Tuffs  2002) in photometric
conditions in 1997 using the Calar Alto 3.5-m telescope in a narrow band
($\Delta\lambda/\lambda = 0.03$) filter centred at 2.26 $\mu$m. The intrinsic
continuum flux density at 2.26 $\mu$m of the nebula is $9.6 \pm 0.7$ Jy, which
is the measured flux density (after star subtraction) of $8.08 \pm 0.03$ Jy,
divided by 0.839 to correct for extinction, following the extinction law found
for stars in the field of the nebula by Wu (1981). The extrapolation to shorter
wavelengths of the Douvion et al.\ power law (scaled to the integrated emission
from the nebula) passes close to the 2.26 $\mu$m point, and the extrapolation
to longer wavelengths intersects with the extrapolation of the radio spectrum
at around 400 $\mu$m.

\begin{figure}%---------------------------------------------------------------%
\centerline{\includegraphics[width=5.4cm,clip=]{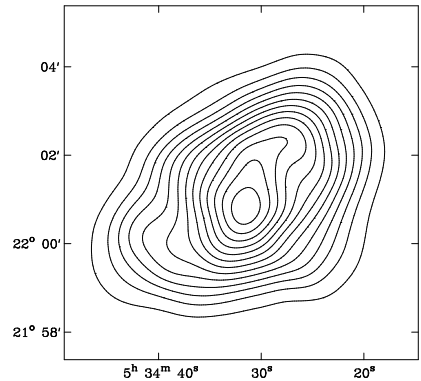}}
\caption{850 $\mu$m image of the Crab nebula smoothed to resolutions of 47
arcsec to match the resolution of the ISOPHOT images at 100 $\mu$m shown in
Fig.~\ref{f:iso60and100}. The contours are linearly spaced.\label{f:smooth}}
\end{figure}%-----------------------------------------------------------------%

\subsection{ISOPHOT 60 and 100~$\mu$m images}\label{s:isoimages}

\begin{figure}%----------------------------------------------------------------%
\centerline{\includegraphics[width=5.4cm,clip=]{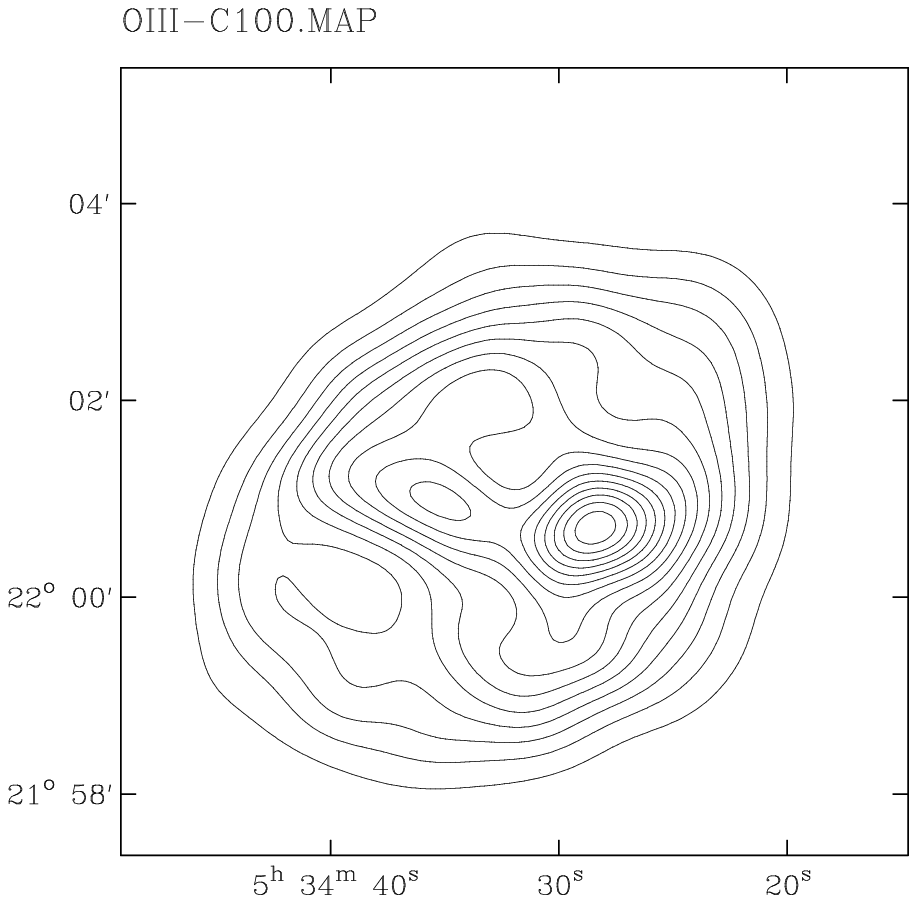}}
\medskip
\centerline{\includegraphics[width=5.4cm,clip=]{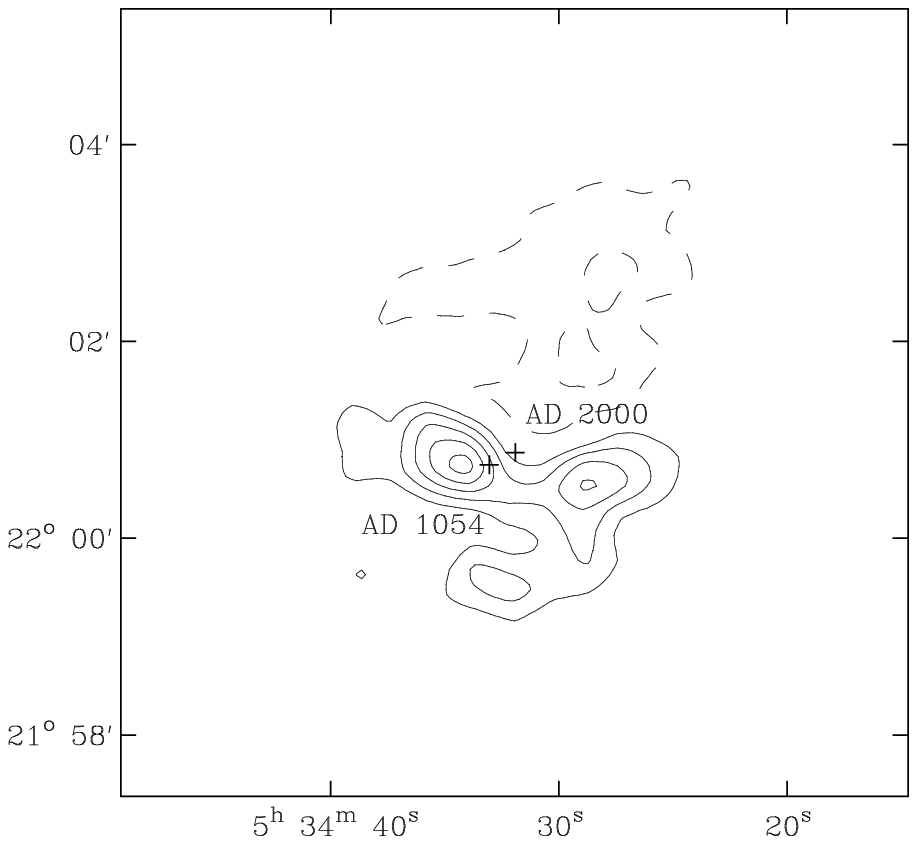}}
\caption{(top) [\OIII] image of the Crab nebula smoothed to resolutions of 47
arcsec to match the resolution of the ISOPHOT image at 100 $\mu$m shown in
Fig.~\ref{f:iso60and100} (bottom) ISOPHOT image of the Crab nebula at 100
$\mu$m, with a scaled version of the 850 $\mu$m removed. The contours
are linearly spaced in each case. The crosses indicate the position
of the Crab pulsar at {\ad} 1054 and {\ad} 2000.\label{f:oxygen}}
\end{figure}

At all three FIR wavelengths the smooth synchrotron emission seen at longer
wavelengths is clearly evident. For comparison, the SCUBA 850-${\mu}$m image
has been convolved to the ISOPHOT beam at 100 ${\mu}$m, see
Fig.~\ref{f:smooth}. However, at 60 and 100 ${\mu}$m two prominent features,
separated by $\approx 80$ arcsec and connected by a ridge running almost
east-west. To a precision of 10 arcsec the positions of the peaks (J2000) are
$05^{\rm h} 34^{\rm m} 28\fs7$, $+22^\circ 00' 30''$ for the western peak and
$05^{\rm h} 34^{\rm m} 34\fs1$, $+22^\circ 00' 45''$ for the eastern peak. The
corresponding offsets from the current position of the pulsar are: 45 arcsec W,
22 arcsec S and 30 arcsec E, 7 arcsec S, respectively, and the corresponding
offsets from Star 16 of Wyckoff \& Murray (1977) are 48 arcsec W, 25 arcsec S
and 27 arcsec E, 10 arcsec S. These features are close to the peaks in the
image of the [\OIII] line emission, from Lawrence et al.\ (1995), convolved to
the resolution of the ISOPHOT 100-${\mu}$m image, which is shown in
Fig.~\ref{f:oxygen}. Within the positional uncertainties, the eastern and
western peaks coincide with absorption spots 3C and 1D, respectively, of Fesen
\& Blair (1990). Part of the eastern peak was imaged by the
HST by Blair et al.\ (1997). The prominent sinuous dust shadow of
extent 25 arcsec seen on the HST images lies within 15 arcsec of the FIR
emission peak which -- like its Western counterpart -- appears slightly
resolved in the ISOPHOT images.
We have scaled the 850 ${\mu}$m image, assuming a power-law extrapolation to
100 ${\mu}$m (with a spectral index of 0.30), and removed this from the
observed 100 ${\mu}$m image. Fig.~\ref{f:oxygen} also shows this synchrotron
removed 100 ${\mu}$m image. The fact that there is evidently less extended
emission in the NW at 100 ${\mu}$m than is expected from a power-law
extrapolation from lower frequencies is consistent with the discussion in
Section~\ref{s:170and850}, that is a spectral turnover between 850 and
170~$\mu$m.

%------------------------------------------------------------------%

\begin{figure}%-----------------------------------------------------------------
\centerline{\includegraphics[angle=0,width=6cm,clip=]{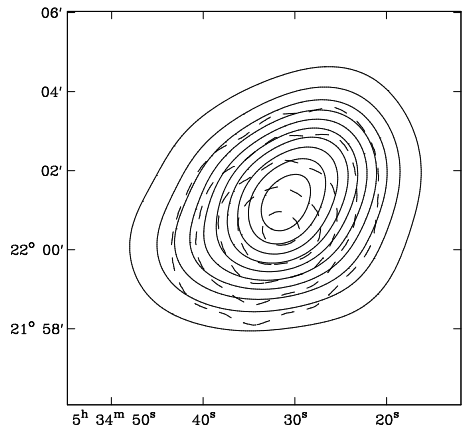}}
\caption{Comparison of the ISOPHOT 170-$\mu$m image of the Crab nebula (dashed
contours), with the JCMT 850-$\mu$m image smoothed to the same resolution
(solid contours). Both sets of contours are linearly spaced.\label{f:850c170}}
\end{figure}%-------------------------------------------------------------------

The correspondence between the 60 and 100 $\mu$m peaks, and the smoothed
[\OIII] emission image, together with the absence of corresponding peaks of
synchrotron emission at 850 ${\mu}$m, suggests that these FIR features are
produced through a thermal emission mechanism. The excess emission could
potentially arise from dust (either in the ejecta or a circumprogenitor
medium), or could be FIR fine structure line emission (also either from ejecta
or from a circumprogenitor medium). In order to determine whether the FIR
excess could be due to line emission, we compared the ISOPHOT surface
brightnesses in the principle lines with the ISOPHOT surface brightnesses seen
at the same positions on maps convolved with the LWS beam -- see
Table~\ref{t:isolws}. We considered the principal lines seen in the spectral
range covered by the C60 and C100 ISOPHOT filters, which are [\OIII] 52
${\mu}$m, [\OI] 63 ${\mu}$m and [\OIII] 88 ${\mu}$m. In all positions at most
11 per cent of the emission seen in the C100 and C60 filters can be attributed
to line emission. In addition, this percentage
seems to be independent of position in the nebula. Therefore, it appears that
the `FIR bump' in the integrated spectra is not due to line emission, and
furthermore that the two compact sources seen on the ISOPHOT 60 and 100 $\mu$m
images are also not due to line emission. Essentially none of the emission seen
in the C160 filter is due to line emission.

\begin{table}
\caption{Estimated contributions to the measured integrated flux density from
synchrotron, line and dust emission at the observed ISO
wavelengths.\label{t:fluxes}}
\begin{tabular}{ccccc}
             & \multicolumn{4}{c}{flux density}            \\ \cline{2-5}
wavelength   & measured  & synchrotron &   line   &  dust  \\
 ($\mu$m)    &   (Jy)    &    (Jy)     &   (Jy)   &   (Jy) \\ \hline
    170      &  $83 \pm 10$ & $105 \pm 9^{\rm a}$ &   --   & $-22 \pm 13$ \\
    100      & $128 \pm 14$ & $ 65 \pm 7^{\rm b}$ &    9   & $ 54 \pm 16$ \\
     60      & $141 \pm 15$ & $ 50 \pm 5^{\rm b}$ &   11   & $ 80 \pm 16$ \\ \hline
\end{tabular}

\medskip\noindent Notes:\\
$^{\rm a}$ extrapolated from the radio spectrum of Baars et al.\ (1977)
corrected for secular changes (see Section \ref{s:spectrum});\\
$^{\rm b}$ extrapolated from the MIR pure synchrotron spectrum of Douvion et
al.\ (see Section \ref{s:spectrum}), with a nominal uncertainty of 10 per cent.
\end{table}

We conclude that the excess emission in the FIR bump (and which is most
prominently in the peaks seen in Fig.~\ref{f:oxygen}) is primarily due to dust
emission. The difference between the relative brightness of these peaks --
i.e.\ at 100 ${\mu}$m, the eastern peak is brighter, whereas at 60 ${\mu}$m,
the western peak is brighter -- would then reflect different dust temperatures
in the peaks, caused by differences in the strength and/or colour of the local
radiation field. To estimate the mass of grains needed to account for the FIR
bump we nevertheless approximate the integrated emission in the bump as arising
from either silicate or graphite grains at a single temperature, with optical
properties given by Laor \& Draine (1993). The FIR dust emission 60 and 100
$\mu$m was calculated -- see Table~\ref{t:fluxes} -- by subtracting from the
observed flux densities the contributions due to line emission (calculated from
an average of the entries in Table~\ref{t:isolws} to be a fraction 0.08 and
0.07 of the total emission in the C60 and C100 filters, respectively) and an
extrapolation of the MIR pure synchrotron power law ($25.9 (\nu/18.7~{\rm
THz})^{-0.5}$ Jy, as derived from the ISOCAM measurements of Douvion et al.\ in
Section~\ref{s:spectrum}). The uncertainties in the derived dust flux densities
are dominated by the uncertainties in the measured flux densities.
Table~\ref{t:fluxes} also includes estimates of the synchrotron and dust
emission at 170 $\mu$m. The dust emission at 60 and 100 ${\mu}$m is consistent
with a small amount of warm dust ($0.01{-}0.07$ M$_{\odot}$ of silicate at
temperatures around 45~K or $0.003{-}0.02$ M$_{\odot}$ of graphite at
temperatures around 50~K). The geometry of this warm dust emission is
consistent with a torus of diameter $\approx 0.8$~pc, presumably created by the
supernova progenitor. There is no evidence for dust emission towards the dark
bays in the optical emission from the remnant in the east and west (Fesen,
Martin \& Shull 1992).

It should be emphasised that the main reason that the FIR bump can be accounted
for by such a moderate quantity of grains is the lack of any evidence for
emission from cold dust in the 170~$\mu$m band of ISOPHOT, which is the most
sensitive of our measured bands for detecting grains with temperatures in the
$10{-}20$~K range. Our nominal estimate for the integrated flux density of the
total dust emission at 170~$\mu$m is $-22 \pm 13$~Jy, a number obtained by
subtracting a synchrotron flux density estimated from an extrapolation of the
long wavelength radio spectrum from the total measured flux density, and
assuming line emission can be neglected in this band (see
Table~\ref{t:isolws}). At 170~$\mu$m, the Rayleigh--Jeans tail of the warm dust
emission detected at 60 and 100~$\mu$m is expected to contribute up to 14~Jy,
implying only $\sim -36 \pm 13$~Jy might be attributable to cold dust, i.e.\ at
nearly the 3$\sigma$ level, our observations are consistent with there being no
cold dust in the nebula. A limit from any cold dust component of 3~Jy at
170~$\mu$m, assuming a grain temperature of 15~K, would correspond to upper
limits of $\sim 0.1$ M$_\odot$ and $\sim 0.4$ M$_\odot$ from cold dust in the
form of graphite and silicate, respectively (for a distance of 2~kpc). These
upper limits are however very crude, since the grain temperature of any cold
dust is not well constrained by our data. Moreover, if the synchrotron spectrum
is turning over at 170 $\mu$m -- as is suggested in Section \ref{s:170and850}
below -- then the simple extrapolation of the synchrotron emission from larger
wavelengths is likely to overestimate the synchrotron emission somewhat. A
constraint on any cold dust arising from the positions of the warm dust
emission peaks is provided by the almost identical morphology seen at
850~$\mu$m and the 20~cm pure synchrotron image (Figs \ref{f:850} and
\ref{f:20}). This suggests that at most $\sim 0.1$~Jy of the 850-$\mu$m
emission can arise from cold dust at each peak, which corresponds to respective
upper limits of $\sim 0.04$ M$_\odot$ and $\sim0.15$ M$_\odot$ of graphite and
silicate assuming a grain temperature of 15~K.

\subsection{Comparison of 170- and 850-$\mu$m images}\label{s:170and850}

The ISOPHOT 170 $\mu$m image shows a centrally brightened structure reminiscent
of that seen at longer wavelengths (i.e.\ synchrotron emission), with no
indication of the peaks seen at 100 and 60 $\mu$m. However, detailed comparison
of the 170 $\mu$m image with the JCMT 850-$\mu$m image, which we take to be
synchrotron emission, does reveal differences in the details of the emission
structure at these two wavelengths. Fig.~\ref{f:850c170} shows the 850-$\mu$m
image smoothed to the resolution of the ISOPHOT 170 $\mu$m image. Comparison
with Fig.~\ref{f:iso160} shows that the peak of the 170-$\mu$m emission (near
$5^{\rm h}$ $34^{\rm m}$ $31^{\rm s}\!\!.7$, $+21^\circ$ $0'$ $30''$), is
displaced by about 40 arcsec to the south of the 850-$\mu$m peak (near $5^{\rm
h}$ $34^{\rm m}$ $30^{\rm s}\!\!.9$, $+21^\circ$ $1'$ $10''$), which is much
larger than any expected uncertainties in the positional accuracy of the
images. Moreover, the 170-$\mu$m emission is relatively fainter in the NW
compared with the expected extrapolation of the synchrotron emission from
850-$\mu$m (see Fig.~\ref{f:oxygen}). One explanation of this, which would also
explain the ISOPHOT 170 $\mu$m flux density being low, is that the effect of a
break in the synchrotron spectrum is becoming evident, but predominantly in the
NW of the remnant, at wavelengths between 850 and 170 $\mu$m. We note that the
optical extent of the Crab nebula agrees well with the radio extent of the
remnant everywhere except in the NW (see Velusamy et al.\ 1984), where the
cm-wavelength radio emission extends beyond the optical emission. Sankrit \&
Hester (1997) note that there is no [\OIII] `skin' in the N and NW where the
radio emission extends beyond the optical. Generally is has long been
recognised (e.g.\ Woltjer 1987) that the extent of the synchrotron emission
decreases with increasing frequency (i.e.\ energy), due to smaller lifetime of
the more energetic particles responsible for the higher energy emission.
However, it is not clear how to relate the deficit of the optical, {\em
thermal}, emission to a spectral turnover at lower frequencies for the {\em
non-thermal} synchrotron emission in the NW.

\subsection{Dust in the remnant}

One of the important results of this investigation is the discovery that
the dust in the Crab Nebula is preferentially associated with the east--west
chain of filaments seen predominantly in [\OIII], lying just south of the
position of the pulsar. This bears some resemblance to the situation
in Cas A, where MIR emission from dust is seen from the [\OIII]-emitting
fast optical filaments (Lagage et al.\ 1996), but contrasts with the situation
in Kepler's SNR, where the MIR emission appears to be associated with
circumstellar dust (Douvion et al.\ 2001). These filaments are also
Helium-rich and have previously been hypothesised to be associated with
circumstellar material responsible for the asymmetry in the north--south
expansion of the nebula (MacAlpine et al.\ 1989; Lawrence et al.\ 1995; Fesen,
Shull \& Hurford 1997). Thus, the dust associated with these filaments may well
also be of circumstellar origin. The ISOPHOT 60 and 100 $\mu$m images reveal
the extent and geometry of the dust-rich circumstellar matter in the nebula,
showing it to be consistent with a torus of diameter $\approx 80$~arcsec (or
0.8~pc for a distance of 2~kpc). This can be compared with the circumstellar
ring around SN~1987A, which has also been shown to be a dust emission source
(Fischera, Tuffs \& V\"olk 2002a), and which has a diameter of 1.3~pc. Unlike
the Crab, which is thought to have had a red supergiant progenitor, the
progenitor of SN~1987A was a blue supergiant star and left no detectable
pulsar-powered nebula behind. The occurrence of quite similar structures
associated with such different progenitors suggests that dusty circumstellar
rings may be commonly associated with supernovae (see also Pozzo et al.\
2004, who show evidence for circumstellar dust around SN 1998S).

About half of the dust emission in the Crab Nebula can be attributed to the
dust torus, and assuming that radiation fields illuminating the grains in the
filaments are not strongly dependent on position within the nebula or on the
chemical composition of the filament, about half of the total dust mass might
be circumstellar in origin. We note however that, as in case of SN~1987A, the
mass of dust in the circumstellar medium measured today could be much lower
than that present prior to {\ad}~1054, due to evaporation of grains by the
ultraviolet-flash from the supernova outburst and subsequent sputtering when
the blast wave reached the circumstellar medium (Fischera, Tuffs \& V\"olk
2002b). This would also modify the grain size distribution by reducing the
abundance of small grains in relation to large grains. Tentative evidence for
large grains sizes in comparison with typical interstellar grains is indeed
provided from a comparison of the near-UV and optical obscuration in HST images
(Blair et al.\ 1997), which shows that the extinction curve is rather flat.

The remaining $\approx 50$ per cent of the dust emission can plausibly be
identified as emanating from supernova condensates in the broadly distributed
filaments distributed over the face of the nebula. However, the total mass of
grains in condensates which can be inferred from our data is at most a few
hundredths of a solar mass. Comparison with the total mass of filaments of $4.6
\pm 1.8$ M$_{\odot}$ (mainly in the form of Helium) in the nebula (Fesen, Shull
\& Hurford 1997) indicates that the total dust-to-gas ratio in the nebula is in
fact at most only of the same order as the interstellar value of $\approx
0.0075$. Thus, even if the condensates seen at the present epoch could
ultimately escape the remnant without being destroyed, the surrounding ISM will
not be significantly enriched in dust. In fact the supernova remnant may have
the opposite effect -- i.e.\ to {\em dilute} the surrounding ISM of grains. We
conclude, therefore, that the progenitors of events like that of {\ad}~1054 are
not significant sources of interstellar dust.

Our observations provide meaningful upper limits on the amount of Carbon that
can be `hidden' in grains in the nebula. The Carbon abundance in the nebula is
of particular importance to our understanding of the progenitor. Carbon
abundances in the gas phase show that the ejecta is Helium rich, but not Carbon
rich, which places an upper limit\footnote{Stars more massive than 8 M$_\odot$
are expected to dredge up freshly synthesised Carbon from the core regions into
the He-rich zone, e.g.\ Nomoto et al.\ (1992).} on the progenitor mass of
$\approx 8$ M$_{\odot}$ when it was on the main sequence (Davidson et al.\
1982). However, this estimate assumes that all the Carbon is in the gas phase.
In the extreme case that all the grains are made of graphite, we infer a
maximum mass of solid state Carbon of $\approx 0.02$ M$_{\odot}$. This can be
compared with the maximum possible amount of Carbon present in the filaments if
they had not been significantly enriched in Carbon produced by nucleosynthesis
in the star. Assuming solar abundances (Anders \& Grevesse 1989) this is a
fraction 0.0022 of the $4.6 \pm 1.8$ M$_{\odot}$ mass of the filaments, which
is $0.010 \pm 0.004$ M$_{\odot}$. This mass is very close to our upper limit on
the Carbon mass of $\approx 0.02$ M$_{\odot}$ which can be in the form of
graphite. We therefore conclude that for the Crab Nebula we have found no
evidence for a substantial boosting of the mass of Carbon in the filaments
through the presence of Carbon in the solid state.

\subsection{Synchrotron Spectral index variations: cm to sub-mm}\label{s:cmtosubmm}

\begin{figure}%---------------------------------------------------------------%
\centerline{\includegraphics[width=8.5cm]{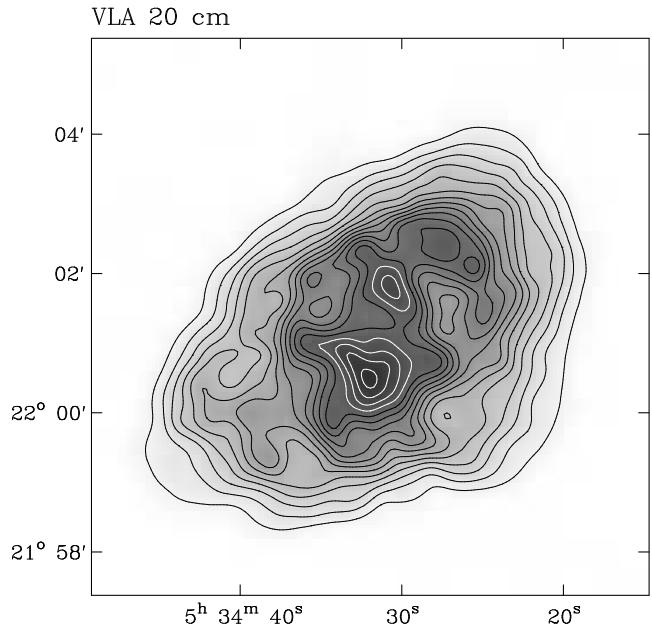}}
\caption{The Crab nebula at 20~cm (1515~MHz) from VLA observations made by
Michael Bietenholz, with a resolution of 17~arcsec for comparison with
Fig.\ref{f:850}. Contours are every 0.4 Jy~beam$^{-1}$.\label{f:20}}
\end{figure}

For a spectral index study of the synchrotron emission from the remnant, the
SCUBA image was compared with a VLA image at 20~cm (1515~MHz), which was kindly
supplied by Michael Bietenholz. The VLA image is made from four arrays of VLA
data observed in 1987 (see Bietenholz \& Kronberg 1991, Bietenholz et al.\
1997). For comparison with the SCUBA image, the VLA image was: (i) scaled by
1.6 per cent in size, as is appropriate for a convergence epoch of {\ad}
1233 for the expansion of the Crab nebula (Bietenholz et al.\ 1991) and the
difference in epochs of the 850-$\mu$m and 20-cm observations, and (ii)
smoothed to a resolution of 17~arcsec to match that of the SCUBA image, see
Fig.\ref{f:20}. (The flux density scale of the VLA image was not changed, as
the expected secular change between 1987 and 1999 is small -- $\approx 2$ per
cent -- compared with the overall uncertainty in the flux density scales of the
observations.) The contour levels in Figs~\ref{f:850} and \ref{f:20} have
been chosen to be at similar relative levels. The 20-cm image is of higher
quality, both in terms of its lower noise and the accuracy of the local
baselevels. The close similarity between these images -- which are at
wavelengths that differ by a factor of {\em over two hundred} -- show that
there is no strong spectral index variation across the remnant. Hence, these
results show that the particle populations responsible for the radio and sub-mm
synchrotron emission are distributed very similarly within the Crab nebula.

%-----------------------------------------------------------------%

\begin{figure}%----------------------------------------------------------------%
\centerline{\includegraphics[width=8.5cm]{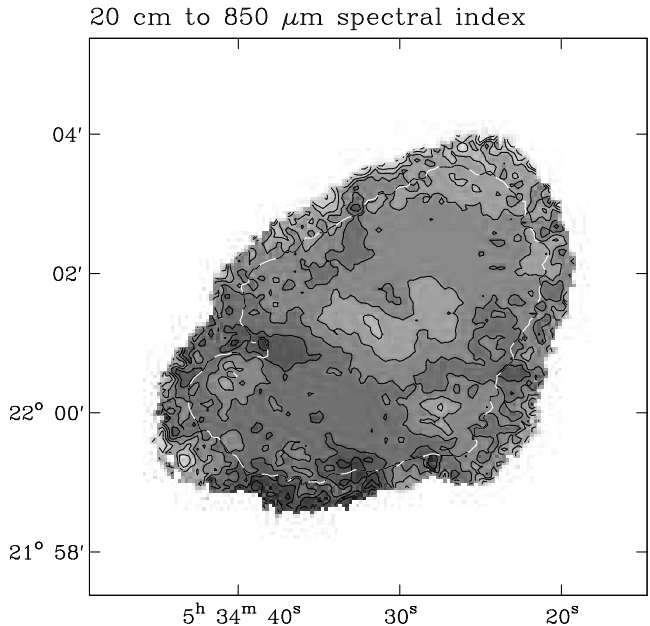}}
\caption{The spectral index, $\alpha$, of the Crab nebula between 20~cm and 850
$\mu$m (see Figs~\ref{f:850} and \ref{f:20}). Contours, and discrete
changes in the shading, are at 0.23 to 0.37 in steps of 0.02 (with the higher
spectral index values being darker). The spectral index is calculated only
where the 850-$\mu$m emission exceeds 0.17 Jy beam$^{-1}$. Also shown is a
single thick, black and white dashed contour of the 850-$\mu$m emission, at
0.43 Jy beam$^{-1}$.\label{f:alpha}}
\end{figure}

An image of the spectral index between 20~cm and 850 $\mu$m, is shown in
Fig.\ref{f:alpha}. The spectral index, $\alpha$, has been calculated where
the 850-$\mu$m emission exceeds 0.17 Jy~beam$^{-1}$. The random uncertainties
in the derived spectral indices are dominated by the uncertainties in the
850-$\mu$m image, not the 20-cm image. Consequently a 850-$\mu$m contour at
$0.43$ Jy beam$^{-1}$ is also shown on Fig.\ref{f:alpha} (i.e.\ at
approximately eight times the variation in the background level of the SCUBA
image, so the uncertainty in the derived spectral index is less than 0.02
inside this contour). These spectral indices are not absolute, as the flux
density scales of the images are not themselves correct in an absolute sense; a
10 per cent systematic shift in either of the flux density scales corresponds
to a constant shift by 0.018 in the derived flux density. The spectral index
over most of the Crab nebula, particularly in regions of the brighter emission,
shows very little variation, with the spectral index being typically between
0.29 and 0.33, with no obvious systematic indication of spectral steepening
towards the edge of the remnant. In comparison with Bandiera et al.'s results,
Fig.~\ref{f:alpha} is less noisy, and also covers a larger area, particularly
to the SW. Bietenholz et al.\ (1997) provide somewhat stronger limits on
spectral variations of the radio emission across the Crab nebula, but over a
much narrower range of frequencies than for our results.

%------------------------------------------------------------------%

There is a slight general gradient of spectral index across the remnant, from
the NW (slightly flatter) to the SE (slightly steeper). This gradient may be an
artefact due either to: (i) a positional offset between the images, or (ii)
different effective sampling of the large scale structure at 850-$\mu$m
compared to 20-cm. It is not thought possible that systematic uncertainties in
the positions of the 850-$\mu$m and 20-cm images could be large enough to
produce spectral gradient seen. But, given the quite different observational
techniques used at 850~$\mu$m and 20~cm, the apparent small spectral index
gradient across the remnant may well be due to differences in effective
sampling of large-scale structures at the two wavelengths. However, if the
spectral gradient is real, the variation may be due to differences in the
spectra of the particles injected into the NW and SE parts of the Crab nebula
from its central pulsar, or may reflect different environments (e.g.\ magnetic
fields) in the NW part of the Crab nebula compared with the SE. In X-rays --
for example from recent Chandra observations (Weisskopf et al.\ 2000) -- there
is also a NW to SE asymmetry (with the NW being brighter), indicating that
there are differences in the energetic relativistic particles and magnetic
fields responsible for the X-ray emission.

The main deviation in spectral index away from the large-scale gradient occur
near the centre of the remnant -- particularly near the position of the pulsar
-- where there are regions with slightly flatter spectra than their
surroundings. The most pronounced region of flatter spectral index is about an
arcmin in maximum extent, elongated NE to SW, centred on the Crab's pulsar (at
$5^{\rm h}$ $34^{\rm m}$ $31.97^{\rm s}$, $22^\circ$ $00'$ $51.9''$), with
another somewhat smaller region of flatter spectrum emission slightly to the
NW. These regions of flatter spectral index were also seen by Bandiera, who
interpreted them -- along with their apparent excess in integrated flux density
at 1.2~mm -- as evidence for a second synchrotron component. Although these
regions of apparent flatter spectrum emission near the centre of the remnant
may be indicative of real spectral variations, it must also be remembered that
images used for the spectral comparison are from different epochs (which also
applies to Bandiera et al.'s comparison). Instead the spectral variations near
the centre of the remnant may reflect temporal variations. Bietenholz
et al.\ (2004) show clear variations in the structure of the radio emission
from the Crab nebula at 5~GHz in an region, $\approx 1.5 \times 1$ arcmin$^2$
in extent around the Crab pulsar, aligned approximately NE to SW. This
corresponds closely with the central region with the largest deviation in
spectral index from the typical values seen in Fig.\ref{f:alpha}.

Other deviations in spectral index, are regions showing flatter spectrum
emission in the SW (near $5^{\rm h}$ $34^{\rm m}$ $27^{\rm s}$, $22^\circ$
$0'$, together with the region just to the NE centre of the remnant (near
$5^{\rm h}$ $34^{\rm m}$ $27^{\rm s}$, $22^\circ$ $1'$ $30''$) -- both also by
Bandiera et al.\ -- and a region of steeper spectrum emission in the W (near
$5^{\rm h}$ $34^{\rm m}$ $40^{\rm s}$, $22^\circ$ $1'$).

Overall there is little variation in spectral index over the face Crab nebula,
over a wide range of frequencies. However, unlike Bandiera et al.\ we do not
see the need to invoke an additional synchrotron component to explain the
emission from the Crab nebula, because: (i) we do not find any obvious excess
in the integrated flux density of the Crab nebula at 850-$\mu$m, and (ii) the
correspondence between the central region of flatter spectrum emission and that
showing temporal variations implies that the apparent flatter spectral index
may well be due to temporal variations.

%==============================================================================%
\section{CONCLUSIONS}\label{s:conclusions}

Here we have presented FIR and sub-mm images of the Crab nebula, which we have
used to investigate both thermal and non-thermal emission from this supernova
remnant. The 60 and 100 $\mu$m ISOPHOT observations show clear excess of
emission above the extrapolated synchrotron radio spectrum at lower frequencies
(as has been see previously from {\sl IRAS} observations). With the improved
angular resolution of the ISOPHOT images, this excess is seen to come
predominantly from two peaks, to the east and west of the centre of the
remnant. From out studies of the LWS spectra we have shown that this excess
indicates the presence of a small amount of warm dust (consistent with
$0.01{-}0.07$ M$_{\odot}$ of silicate at temperatures around 45~K or
$0.003{-}0.02$ M$_{\odot}$ of graphite at temperatures around 50~K) in the
remnant. The distribution of this dust is consistent with a torus of diameter
$\approx 0.8$~pc created by the supernova progenitor prior to its explosion,
superimposed upon a broadly  distributed component which may be supernova
condensates in the filaments. Since the dust-to-gas ratio in the filaments is
comparable to the interstellar value, even if the condensates seen at the
present epoch could ultimately escape the remnant without being destroyed, the
surrounding ISM will not be significantly enriched in dust. Our upper limit of
$\approx 0.02$ M$_{\odot}$ on the total mass of Carbon in the form of graphite
is consistent with the inference from the gas-phase Carbon abundances that
there has been no significant synthesis of Carbon in the progenitor during its
lifetime. The lower resolution 170 $\mu$m ISOPHOT image does not show any
excess emission, but instead is possibly fainter, particularly in the NW, than
expected from an extrapolation of the lower-frequency radio synchrotron
spectrum. These, and other observations, are consistent with the synchrotron
spectrum becoming gradually steeper throughout the FIR spectral range covered
by the ISOPHOT observations. The flux density of the Crab nebula from SCUBA
850-$\mu$m observations is consistent with an extrapolation of radio
synchrotron spectrum from lower frequencies. Comparison of the 850-$\mu$m image
with a 20-cm VLA image shows there is little variation in spectral index across
the face of the remnant between these wavelengths. Although there are some
spectral variations near the centre of the remnant, as has been seen
previously, we do not see the need for the second radio synchrotron component,
such as that proposed by Bandiera et al.

%==============================================================================%
\section*{Acknowledgements}

We are grateful to S.\ Lawrence and M.\ Bietenholz for making available their
[\OIII] and 20-cm images of the nebula in [\OIII], and to D.\ Pierce-Price and
J.\ Richer and for help and advice with the MEM processing of the JCMT data.
The ISOPHOT data presented in this paper were reduced using P32Tools, a PHT32
processing and transient correction program developed at the
Max-Planck-Institut f\"ur Kernphysik, and incorporated into PIA (the
interactive analysis package for ISOPHOT) by the ISO Data Centre of the ESA
Research and Scientific Support Department in collaboration with the Infrared
Processing and Analysis Center (IPAC). PIA is a joint development by the ESA
Astrophysics Division and the ISOPHOT Consortium with the collaboration of
IPAC. Contributing ISOPHOT consortium institutes are the Dublin Institute for
Advances Studies, the Rutherford Appleton Laboratory, the Astrophysics
Institute Potsdam, the Max-Planck-Institut f\"ur Kernphysik, and the
Max-Planck-Institut f\"ur Astronomy. The JCMT is operated by the Joint
Astronomy Centre in Hilo, Hawaii on behalf of the parent organisations Particle
Physics and Astronomy Research Council in the United Kingdom, the National
Research Council of Canada and The Netherlands Organization for Scientific
Research.

%==============================================================================%

\end{document}